\documentclass[
aps,
prl,
showpacs,
floatfix,
reprint,
twoside,
superscriptaddress
]{revtex4-1}
\usepackage{amsmath}
\usepackage[T1]{fontenc}
\usepackage{fourier}
\usepackage{graphicx}
\usepackage[colorlinks=true,
allcolors=blue,
dvipdfm=true]{hyperref}

\begin{document}

\title{Amplitude Higgs mode and admittance in superconductors with a moving condensate}

\author{Andreas Moor}
\affiliation{Theoretische Physik III, Ruhr-Universit\"{a}t Bochum, D-44780 Bochum, Germany}
\author{Anatoly F.~Volkov}
\affiliation{Theoretische Physik III, Ruhr-Universit\"{a}t Bochum, D-44780 Bochum, Germany}
\author{Konstantin B.~Efetov}
\affiliation{Theoretische Physik III, Ruhr-Universit\"{a}t Bochum, D-44780 Bochum, Germany}

\begin{abstract}
We consider the amplitude (Higgs) mode in a superconductor with a condensate flow (supercurrent). We demonstrate that, in this case, the amplitude mode corresponding to oscillations~$\delta |\Delta|_{\Omega} \exp(i \Omega t)$ of the superconducting gap is excited by an external ac electric field~$\mathbf{E}_{\Omega} \exp(i \Omega t)$ already in the first order in~$|\mathbf{E}_{\Omega}|$, so that ${\delta |\Delta|_{\Omega} \propto (\mathbf{v}_{0} \mathbf{E}_{\Omega})}$, where~$\mathbf{v}_{0}$ is the velocity of the condensate. The frequency dependence~$\delta |\Delta|_{\Omega}$ has a resonance shape with a maximum at ${\Omega = 2 \Delta}$. In contrast to the standard situation without the condensate flow, the oscillations of the amplitude~$\delta |\Delta(t)|$ contribute to the admittance~$Y_{\Omega}$. We provide a formula for admittance of a superconductor with a supercurrent. The predicted effect opens new ways of experimental investigation of the amplitude mode in superconductors and materials with superconductivity competing with other states.
\end{abstract}

\date{\today}

\maketitle

Recent development of terahertz technology (see for a review Refs.~\cite{terahertz,Basov_et_al_2011}) has made it possible to systematically investigate the amplitude mode (AM) in superconductors~\cite{ExpS13,Matsunaga_et_al_2014,Beck_et_al_2013}. The AM in the superconductors resembles gapful Higgs modes in field theories that can be interpreted as Higgs bosons~\cite{Higgs64}. The similarity of quantum field theory and cosmology to superconductivity and other ordered phases in condensed matter has intensively been discussed previously~\cite{Shifman_Yung_2003,Vilenkin,Volovik13,Katsimiga_2015} and the attempts to probe the AM in superconductors was stimulated to a large extent by this similarity.

The superconducting AM is gapful with a comparatively large gap~$\Delta $ and hence high frequencies ${\Omega \sim \Delta}$ are needed. Moreover, its observation demands a rather sophisticated technique of femtosecond optical pump-probe spectroscopy developed only in the last decades. Therefore, it is of no surprise that the AM has not been identified experimentally earlier.

The AM mode describing variations of the modulus of the order parameter differs from the well known phase collective mode~(CM) in superconductors~\cite{Carlson_Goldman_1975,Schmid_Schoen_1975,ArtVolkov75} and, in contrast to it, is not accompanied by perturbations of the charge density.

A collisionless relaxation of a small perturbation of the energy gap~$\delta |\Delta (t)|$ has been described in Ref.~\cite{VolkovKogan73} where it has been shown that it oscillates and decays in time in a power law fashion,
\begin{equation}
\delta |\Delta (t)| \sim \delta |\Delta (0)| \frac{\cos (2\Delta_{0} t)}{\sqrt{2 \Delta_{0} t}} \,, \label{a2}
\end{equation}
where~$\Delta_0$ is the unperturbed superconducting order parameter. Equation~(\ref{a2}) has only recently been confirmed experimentally~\cite{ExpS13,Matsunaga_et_al_2014}. Nonlinear solutions for the time dependence of the perturbation~$\delta |\Delta (t)|$ in superconductors have been published in the last decade~\cite{Levitov04,*Levitov04a,Amin_et_al_2004,Simons05,Warner_Leggett_2005,Altshuler05,Altshuler05a,Yuzbashyan06,Gurarie09,Foster_et_al_2013,Yuzbashyan_et_al_2015}.

Various aspects of the AM and methods of its detection have been considered in recent publications. Probing the AM by measuring time-dependent photoemission spectra has been suggested in Ref.~\cite{Kemper_et_al_2015} for external perturbations of different strength. Nonlinear absorption of ac electromagnetic field in a superconductor (third harmonic generation and two-photon absorption) and corresponding excitation of the AM has been studied in Refs.~\cite{Aoki15,Cea15,Jujo15,Cea_et_al_2016,Murakami_Werner_Tsuji_Aoki_2016}. The AM in superconductors with a strong electron-phonon coupling has been studied in recent papers Refs.~\cite{Manske14,Kemper16} and the AM in $d$\nobreakdash-wave superconductors has been analyzed in Ref.~\cite{Peronaci15}.

In all the previous theoretical papers, the ac electric field~$\mathbf{E}_{\Omega}$ acting on a superconductor (for instance,~$\mathbf{E}_{\Omega}$ in a laser pulse) is assumed to be sufficiently strong, so that the second order~$|\mathbf{E}_{\Omega}|^{2}$ is sufficiently large. This requirement is due to the fact that only the second order (or higher even orders) of the electric field~$\mathbf{E}_{\Omega}$ can couple to the perturbation~$\delta |\Delta|$, which is natural because~$|\Delta|$ is a scalar whereas~$\mathbf{E}_{\Omega}$ is a vector. Action of a short laser pulse on a superconductor used in experiments Refs.~\cite{ExpS13,Matsunaga_et_al_2014} destroys Cooper pairs leading to sudden suppression of the order parameter~$\Delta$. After the end of the laser pulse the perturbation~$\delta |\Delta |$ relaxes oscillating in time in accordance with Eq.~(\ref{a2}). This evolution of~$\delta |\Delta (t)|$ is traced with the help of an additional weak probe pulse whose transmission or reflection coefficients depend on the instant magnitude of~$\delta |\Delta (t)|$.

\begin{figure}[tbp]
\includegraphics[width=1.0\columnwidth]{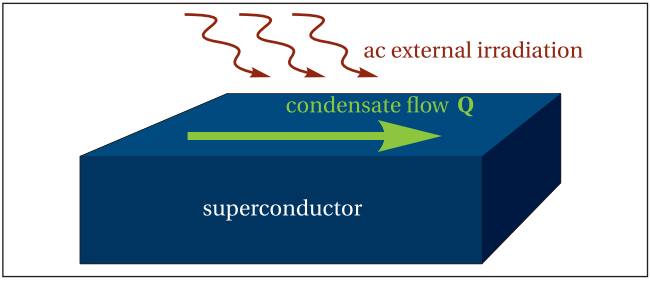}
\caption{(Color online.) Schematic representation of the system under consideration.}
\label{fig:System}
\end{figure}

\begin{figure}[tbp]
\includegraphics[width=1.0\columnwidth]{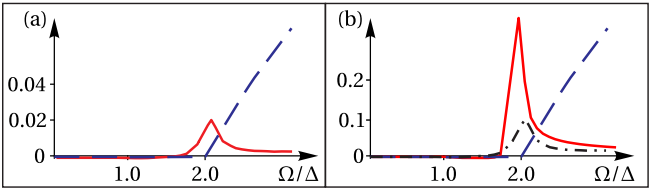}
\caption{(Color online.) The frequency dependence of the real part of the admittance normalised to its value in the normal state and corresponding to different parts of the ac currents~$\mathbf{I}_{\Omega}$. The dashed line~\cite{Mattis_Bardeen_1958,AbrGor58,tinkham2004introduction} corresponds to the real part of the first term in Eq.~(\ref{a21}) [described by~$I_0^{(1)}$ in Eq.~(\ref{A3a}) of Supplemental Material]. The peak in panel~(a) corresponds to the second term in Eq.~(\ref{a21}) [described by~$\delta I^{(1)}$ in Eq.~(\ref{A4}) of Supplemental Material]. The most important peak in panel~(b) corresponds to the third term in Eq.~(\ref{a21}) [described by~$\delta I^{(2)}$ in Eq.~(\ref{A6}) of Supplemental Material]. Note that the scale in the panel~(a) differs from that of the panel~(b). The dash-dotted black line in panel~(b) represents the line shown in panel~(a)---multiplied by a factor of~$5$ to be visible in the plot.}
\label{fig:3}
\end{figure}

In this paper, we consider the AM in a superconductor in the presence of a condensate flow with momentum~$\mathbf{Q}_{0}$ as sketched in Fig.~\ref{fig:System}. It will be shown that, in this case, the mechanism of the AM excitation is quite different---it is excited by a weak ac electric field~$\mathbf{E}_{\Omega}$ which induces ac condensate momentum~$\mathbf{Q}_{\Omega}$ so that the amplitude of the AM~${\delta |\Delta_{\Omega}| \sim (\mathbf{Q}_{\Omega} \mathbf{Q}_{0})}$ is linear in the field~$E_{\Omega}$. Moreover, the AM contributes to the admittance of the superconductor~$Y(\Omega)$ leading to a sharp peak in $\mathrm{Re} [Y(\Omega)]$ at the frequency ${\Omega \simeq 2 \Delta}$, see Fig.~\ref{fig:3}. The effect of the AM on the admittance of superconductors with moving condensate is novel, although attempts to calculate the impedance of in this situation have already been undertaken~\cite{Garfunkel_1968,Semenov_et_al_2016,Gurevich_2014,Clem_Kogan_2012}. In other words, the AM can be probed in the presence of the supercurrent already by measuring the impedance at frequencies~${\Omega \simeq 2\Delta}$. At a fixed frequency~$\Omega $, one can reach a resonance behavior in the vicinity of ${\Omega = 2\Delta(T)}$ by varying the temperature~$T$. It is important to note that the contribution of the AM to the impedance is zero if the polarization of the incident electromagnetic wave is perpendicular to the direction of the vector~$\mathbf{Q}_{0}$. No doubt, realizing the proposed effect experimentally will lead to a considerably better understanding of properties of the AM not only in conventional BCS superconductors, but also in high\nobreakdash-$T_{\text{c}}$ superconductors with coexisting order parameters since the additional (not superconducting) OP is not affected by a present condensate flow.

Although explicit calculations leading to this result are rather involved, the main reason for this unusual behavior can rather easily be understood. The supercurrent is characterized by condensate velocity ${\mathbf{v}_{\text{S}} = \mathbf{Q} (t) / m}$, where
\begin{equation}
\mathbf{Q}(t) = [\mathbf{\nabla} \chi - 2 \pi \mathbf{A}(t) / \Phi_{0}] / 2 \label{a3}
\end{equation}
is the gauge-invariant condensate momentum, $\chi$~is the phase of the order parameter~$\Delta$, $\mathbf{A}(t)$ is the vector potential, ${\Phi_{0} = c h / ( 2e )}$ is the magnetic flux quantum, and~$m$ is the electron mass. The condensate momentum~$\mathbf{Q}(t)$ determines the interaction between the electric field and the modulus~$|\Delta|$ of the superconducting order parameter. Using the gauge invariance we write the corresponding term~$S_{\text{int}}$ in the action in the standard form
\begin{equation}
S_{\text{int}} = \int C \mathbf{Q}^{2}(t) |\Delta(t)|^{2} \, dt \, d\mathbf{r} \,, \label{a4}
\end{equation}
where~$C$ is a constant and~$\mathbf{Q}(t)$ can be written as
\begin{equation}
\mathbf{Q}(t) = \mathbf{Q}_{0} + \mathbf{Q}_{\Omega}(t) \,, \label{a5}
\end{equation}
where $\hbar \mathbf{Q}_{0}=\mathbf{v}_{0}/m$, and~$\mathbf{v}_{0}$ is the velocity corresponding to the dc current~$\mathbf{I}_{0}$. The time dependent part ${\mathbf{Q}_{\Omega}(t) = \mathrm{Re}\big[\mathbf{Q}_{\Omega} \exp(i \Omega t)\big]}$ of the momentum is proportional to the incident electric field ${\mathbf{E}_{\Omega}(t) = \mathrm{Re}\big[\mathbf{E}_{\Omega} \exp(i \Omega t)\big]}$,
\begin{equation}
\mathbf{E}_{\Omega} = i \Omega (\hbar / em) \mathbf{Q}_{\Omega} \,. \label{a6}
\end{equation}

Writing the time dependence of the absolute value~$|\Delta(t)|$ as
\begin{equation}
|\Delta(t)| = \bar{\Delta} + \mathrm{Re}\big[ \delta |\Delta|_{\Omega} \exp(i \Omega t) + \delta |\Delta|_{2\Omega} \exp(2i \Omega t) \big] \,, \label{a7}
\end{equation}
we reduce the action~$S_{\text{int}}$ to the form
\begin{align}
S_{\text{int}} = S_{0} &+ 4 C \mathrm{Re} \int \delta |\Delta|_{\Omega} |\bar{\Delta}| \mathbf{Q}_{0} \mathbf{Q}_{-\Omega} \, d\mathbf{r} \label{a8} \\
                &+ C \mathrm{Re} \int \big[ 2 \delta |\Delta|_{2 \Omega} |\bar{\Delta}| + \big( \delta |\Delta|_{\Omega} \big)^{2} \big] \mathbf{Q}_{-\Omega}^{2} \, d\mathbf{r} \,, \notag
\end{align}
where~$S_{0}$ does not contain~$\mathbf{Q}_{\Omega}$.

In the absence of the dc current~$\mathbf{I}_{0}$, the second term in the first line of Eq.~(\ref{a8}) vanishes and the action contains only the quadratic in the electric field terms written in the second line. This is the standard situation and the experiments Refs.~\cite{ExpS13,Matsunaga_et_al_2014,Beck_et_al_2013} used this type of the coupling to the laser field for probing the AM.

However, the finite dc supercurrent~$\mathbf{I}_{0}$ makes the linear coupling of the electric field~$\mathbf{E}(t)$ to the AM possible and the second term in Eq.~(\ref{a8}) describes this coupling, which leads to oscillating perturbations of the gap. It is interesting to note that in both the cases the AM does not lead to density oscillations and the possibility of the excitation of this mode by~$\mathbf{E}(t)$ in the linear approximation is not related to the charge oscillations. Below, we concentrate on studying the linear response to the electric field~$\mathbf{E}(t)$.

Of course, the presented heuristic arguments are not sufficient for deriving final formulas and we make explicit calculations using the formalism of quasiclassical Green's functions (see Supplemental Material). We present first the final results in a form that can easily be understood without going into details.

\begin{figure}[tbp]
\includegraphics[width=1.0\columnwidth]{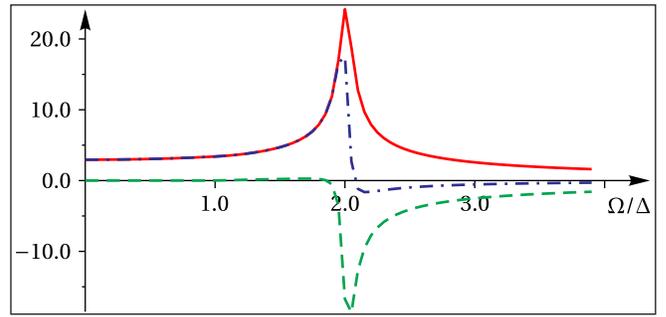}
\caption{(Color online.)  Dependence of~$\frac{\delta \Delta}{i \delta W_{Q}}$ on~$\Omega$ [see Eq.~(\ref{17})]. As seen, $\big|\frac{\delta \Delta}{i \delta W_{Q}}\big|$ shows a peak (resonance) as function of~$\Omega$ at ${\Omega = 2 \Delta}$ (solid red line). Also, $\mathrm{Re}\big(\frac{\delta \Delta}{i \delta W_{Q}}\big)$ (dashed green line) and $\mathrm{Im}\big(\frac{\delta \Delta}{i \delta W_{Q}}\big)$ (dash-dotted blue line) are displayed. We set ${\gamma = 0.05 \Delta}$ and the temperature ${T = 0.05 \Delta}$.}
\label{fig:2}
\end{figure}

We have found that the oscillating electric field~$\mathbf{E}(t)$ incident onto a superconducting moving condensate with the momentum~$\mathbf{Q}_{0}$ leads to an oscillating perturbation~$\delta |\Delta|_{\Omega} \exp(i \Omega t)$ of the superconducting order parameter with the amplitude that can be written as
\begin{equation}
\delta |\Delta|_{\Omega} = D (\mathbf{Q}_{0} \mathbf{Q}_{\Omega}) F(\Omega) \,,  \label{a9}
\end{equation}
where~$D$ is the diffusion coefficient and the momenta~$\mathbf{Q}_{0,\Omega}$ are given by Eq.~(\ref{a5}). The function~$F(\Omega)$ depends on the frequency of the ac field~$\Omega$. Its explicit form is given in Eq.~(\ref{17}) and presented in Fig.~\ref{fig:2} demonstrating a resonance at ${\Omega = 2 \Delta}$. This is in agreement with the observation of the free oscillations of~$\delta |\Delta (t)|$ [see Eq.~(\ref{a2})] caused by a laser pulse~\cite{ExpS13}, but contrasts a resonance at ${\Omega = \Delta}$ found in Ref.~\cite{Matsunaga_et_al_2014}. The latter observation was due to two-photon absorption caused by intensive pump laser pulse in the absence of dc supercurrent. The second weak probe pulse served as a tool to trace the temporal evolution of~$\delta |\Delta(t)|$. Although one can probe the AM with the aid of similar optic methods, the linear dependence of the current on the electric field obtained here allows one to detect the AM simply by measurements of the impedance of the system~$Z(\Omega)$ as a function of the dc current~$I_{0}$ and the frequency. The impedance can be extracted from the coefficients of reflection or transmission of one pulse of the light irradiating the superconductor with a supercurrent. We have found that the current~$I_{\Omega}$ contains, in particular, the terms ${I_{\Omega} \sim \delta \Delta_{\Omega}}$ showing the resonance behavior (see Supplemental Material). In this case, the current can be written in the form
\begin{equation}
\mathbf{I}_{\Omega} = K(Q_{0},T) \mathbf{Q}_{\Omega} + \varkappa_{\text{res}}(\Omega) \big(\mathbf{Q}_{\Omega} \mathbf{Q}_{0}\big) \mathbf{Q}_{0} \,, \label{19}
\end{equation}
where~$\varkappa_{\text{res}}(\Omega)$ is a function demonstrating a resonance at ${\Omega = 2 \Delta}$ and defined in Eq.~(\ref{A10}).

Although the first term in Eq.~(\ref{19}) has a standard form of the response to an external electric field, the second one demonstrates the new effect of the excitation of the~AM and is the main result of this paper. The projection~$I_{\parallel \Omega}$ of the current~$\mathbf{I}_{\Omega}$ on the $\mathbf{E}_{\Omega}$~direction determines the admittance ${Y(\Omega) = I_{\parallel \Omega} / E_{\Omega}}$ and we obtain for this quantity in the limit of small~$Q_{0}$
\begin{equation}
Y(\Omega) = \frac{e}{i \Omega} \Big[ K(0,\Omega) + Q_{0}^{2} \frac{\partial K(Q_{0},\Omega)}{\partial Q_{0}^{2}} + \varkappa_{\text{res}}(\Omega) Q_{0}^{2} \cos^{2} \vartheta \Big] \,,
\label{a21}
\end{equation}
where~$\vartheta$ is the angle between the~${\mathbf{E}_{\Omega} || \mathbf{Q}_{\Omega}}$ and~$\mathbf{v}_{0}$ vectors. In Eq.~(\ref{a21}), the first term in the brackets stands for the linear response, which has been found at ${Q_{0} = 0}$ by Mattis and Bardeen~\cite{Mattis_Bardeen_1958} and Abrikosov and Gor'kov~\cite{AbrGor58}, and the second one is a correction to the linear response due to the moving condensate (this term has been analyzed for arbitrary~$\mathbf{Q}_{0}$ and small~$\Omega$ in Ref.~\cite{Clem_Kogan_2012}).

The third term in Eq.~(\ref{a21}) was overlooked in all the previous studies of the superconductors with moving condensate, Refs.~\onlinecite{Semenov_et_al_2016,Gurevich_2014,Clem_Kogan_2012}. Actually, it is the resonant term describing the excitation of the AM. It strongly depends on the angle~$\vartheta$ turning to zero for perpendicular polarization of the vectors~$\mathbf{Q}_{0}$ and~$\mathbf{Q}_{\Omega}$. This dependence enables a simple method of experimental separation between the conventional contributions and the new one corresponding to the excitation of the~AM.

The frequency dependence of the admittance~$Y(\Omega)$~(see Supplemental Material) is represented in Fig.~\ref{fig:3}.

Now, we turn to a systematic calculation of the response of the AM to the electric field. We consider a BCS superconductor in the diffusive limit in the presence of the condensate flow and ac external irradiation. We assume that all quantities are uniform in space. This condition can be achieved in a thin superconducting film with a thickness less than the London penetration and skin depth. The dynamics of the order parameter~$\Delta$ is described by the Usadel equation~\cite{Usadel} generalized for a non-equilibrium case~\cite{LO,BelzigRev,RammerSmith,Kopnin},
\begin{equation}
\epsilon \check{\tau}_{3} \check{g} - \check{g} \check{\tau}_{3} \epsilon^{\prime} + \big[\check{\Delta} \,, \check{g}\big] - i D \nabla \big(\check{g} \nabla \check{g}\big) = 0 \,,
\label{1}
\end{equation}
where~$\check{g}(\epsilon,\epsilon^{\prime})$ is a matrix Green's function defined as a Fourier transform of a two-times Green's function.

The diagonal elements of the matrix~$\check{g}$ are the retarded (advanced) Green's functions~$\hat{g}^{R/A}$, and the off-diagonal element~$\check{g}|_{12}$ is the Keldysh Green's function~$\hat{g}^{K}$~\cite{Keldysh_1965}. The matrices~$\check{\tau}_{3}$ and~$\check{\Delta}$ are diagonal matrices with elements~$\hat{\tau}_{3}$ and~$\hat{\Delta}$. The superconducting order parameter ${\hat{\Delta} = \Delta (i \hat{\tau}_{2} \cos \chi + i \hat{\tau}_{1} \sin \chi)}$ depends on the phase~$\chi$.

Making the gauge transformation
\begin{equation}
\check{g}(t,t^{\prime}) = \check{S}(t) \cdot \check{g}_{\text{n}}(t,t^{\prime}) \cdot \check{S}^{\dagger}(t^{\prime}) \,, \label{3}
\end{equation}
where the matrix~$\check{S}(t)$ is a diagonal matrix with the elements ${\hat{S} = \exp (i \hat{\tau}_{3} \chi /2)}$, we bring Eq.~(\ref{1}) to the form (the subscript ``n'' is omitted)
\begin{equation}
\epsilon \check{\tau}_{3} \check{g} - \check{g} \check{\tau}_{3} \epsilon^{\prime} + \big[\check{\Delta} \,, \check{g}\big] = -iD \big[\mathbf{Q} \check{\tau}_{3} \,, \check{g} \big[\mathbf{Q} \check{\tau}_{3} \,, \check{g}\big]\big] \,.
\label{4}
\end{equation}
Equation~(\ref{4}) is supplemented by the normalization condition
\begin{equation}
\check{g} \cdot \check{g} = 1 \,.
\label{a11}
\end{equation}

Solving the non-linear equation~(\ref{4}) with the constraint~(\ref{a11}) is generally not an easy task. However, the solution can comparatively easily be found in the linear approximation in the irradiation field~$\mathbf{E}_{\Omega}$ entering only the RHS of this equation. We should also take into account that the value of the gap is reduced in the presence of the condensate flow but we consider this reduction also as a small perturbation. In order to justify these approximations we assume that both~$Q_{0}$ and~$Q_{\Omega}$ are small, i.e., ${D Q_{0,\Omega}^{2} \ll \Delta}$. Then, we have to find the response of the superconductor to finite~$\mathbf{Q}_{0,\Omega}$ considering the RHS of Eq.~(\ref{4}) as a small perturbation.

In the zeroth approximation, the RHS of Eq.~(\ref{4}) vanishes and the elements of the equilibrium matrix~$\check{g}_{0}$ containing on the diagonal the retarded~$\hat{g}_{0}^{R}$, the advanced~$\hat{g}_{0}^{A}$, and the Keldysh Green's functions~$\hat{g}_{0}^{K}$ as the $12$ element, are well known
\begin{align}
\hat{g}_{0}^{R(A)} &= g_{0}^{R(A)} \hat{\tau}_{3} + i \hat{\tau}_{2} f_{0}^{R(A)} \,, \\
\hat{g}_{\text{st}}^{K} &= (\hat{g}_{0}^{R} - \hat{g}_{0}^{A}) \tanh (\epsilon / 2T) \,,
\label{a12}
\end{align}
where
\begin{equation}
g_{0}^{R(A)}(\epsilon) = f_{0}^{R(A)}(\epsilon) \Delta / \epsilon = \epsilon / \zeta_{0}^{R(A)}(\epsilon) \,,
\end{equation}
and ${\zeta_{0}^{R(A)}(\epsilon) = \sqrt{(\epsilon \pm i \gamma)^{2} - \Delta_{0}^{2}}}$. The constant ${\gamma \rightarrow + 0}$ enables choosing the proper branch of the square root although a finite value of~$\gamma$ can be related to different sources.

The RHS in Eq.~(\ref{4}) contains two characteristic energies, ${D Q_{0}^{2}}$ and ${D \mathbf{Q}_{0} \mathbf{Q}_{\Omega}}$, which are assumed to be small compared to~$\Delta$. Writing
\begin{align}
\check{g} &= \check{g}_{0} + \delta \check{g}_{0} + \delta \check{g}_{\Omega} \,, \\
\Delta &= \Delta_{0} + \delta \Delta _{0} + \delta \Delta _{\Omega} \,,
\label{a14}
\end{align}
where~$\delta \check{g}_{0}$ and~$\delta \Delta_{0}$ are proportional to~$Q_{0}^{2}$, while~$\delta \check{g}_{\Omega}$ and $\delta \Delta_{\Omega}$ are proportional to~$Q_{0}^{2} Q_{\Omega}$, we reduce Eqs.~(\ref{a11}),~(\ref{4}) and the corresponding self-consistency equation for the order parameter to linear equations for~$\delta \check{g}_{0}$, $\delta \Delta_{0}$, $\delta \check{g}_{\Omega}$, and $\delta \Delta_{\Omega}$.

Here, we display only the final analytical expression for the oscillating part~$\delta \Delta_{\Omega}$ of the order parameter. The result obtained for arbitrary temperature can be written in the form
\begin{equation}
\delta \Delta_{\Omega} = \frac{i \delta W_{Q} \big[ B_{\Omega}^{R} (\Omega, \Delta_{0}) - B_{\Omega}^{A}(\Omega, \Delta_{0}) + B_{\Omega}^{\text{an}}(\Omega, \Delta_{0}) \big]}{A_{\Omega}^{R}(\Omega, \Delta_{0}) - A_{\Omega}^{A}(\Omega, \Delta_{0}) + A_{\Omega}^{\text{an}}(\Omega, \Delta_{0})} \,, \label{17}
\end{equation}
where ${\delta W_{Q} = D \mathbf{Q}_{0} \mathbf{Q}_{\Omega}}$.

In Eq.~(\ref{17}),~$A_{\Omega}^{R(A)}$, $A_{\Omega}^{\text{an}}$, $B_{\Omega}^{R(A)}$, and~$B_{\Omega}^{\text{an}}$ are functions of temperature~$T$ and frequency~$\Omega$~(see Supplemental Material). Equation~(\ref{17}) describes a correction to the superconducting order parameter due to the linear coupling of electromagnetic field to the modulus of the order parameter. This contribution was not considered so far.

Note that the denominator in Eq.~(\ref{17}) is close to zero at ${\Omega \simeq 2 \Delta}$, which determines the resonance frequency of the AM (Higgs mode). The frequency dependence of the function~$\delta \Delta_{\Omega}$ is depicted in Fig.~\ref{fig:2}. One can see the resonance at~${\Omega = 2\Delta}$, which is a very important feature of the frequency dependence of the perturbation of the superconducting gap.

Having found the corrections~$\delta \check{g}_{\Omega}$ and~$\delta \Delta_{\Omega}$, we can calculate~(see Supplemental Material) the admittance~$Y(\Omega)$, Eq.~(\ref{a21}). In particular, we are interested in the third term, which is related to the excitation of the AM and leads to a sharp peak in $\mathrm{Re}[Y(\Omega)]$ at ${\Omega = 2 \Delta}$. This term is larger than the second one at low frequencies ${\Omega \ll \Delta}$. The admittance~$Y(\Omega)$ can be extracted from the measurements of reflection of the light irradiating a superconductor with a supercurrent~(see Supplemental Material). One can estimate the normalized conductance ${\tilde{\sigma}(\Omega) = \mathrm{Re}[Y(\Omega)] / \mathrm{Re}[Y_{N}(\Omega)]}$, where~$Y_{N}(\Omega)$ is the admittance in the normal state. For the most important third term in Eq.~(\ref{a21}) we obtain ${\tilde{\sigma}(\Omega) \simeq (D Q_{0}^{2} / \Delta) |\delta \Delta _{\Omega}| / \Delta \simeq (Q_{0} / Q_{\text{cr}})^{2} |\delta \Delta_{\Omega} | / \Delta}$, where~$Q_{\text{cr}}$ is the critical momentum of the moving condensate (${Q_{\text{cr}}^{2} \simeq \Delta /D}$). Taking for estimates ${\gamma \simeq D Q_{0}^{2}}$, (this approximation qualitatively describes the smearing of the BCS density-of-states due to moving Cooper pairs~\cite{Volkov199321,Anthore_Pothier_Esteve_2003,Clem_Kogan_2012}), we obtain at the resonance point ${\tilde{\sigma}( 2 \Delta) \approx 1}$. This means that the height of the peak is of the order of the conductance in the normal state and can be measured. The frequency corresponding, for example, to~$\Delta$ in~Al $({T_{\text{c}} \approx 1.2~\text{K}}$, ${\Delta = 177 \mu\text{eV}}$) is of the order of~$50$~GHz~\cite{de_Visser_et_al_2014}. In the case of high\nobreakdash-$T_{\text{c}}$ superconductors, the characteristic frequencies are shifted to THz~frequency range. Note that the dashed line in Fig.~\ref{fig:3} corresponds to the absorption coefficient in superconductors in absence of a condensate flow~\cite{Mattis_Bardeen_1958,AbrGor58} measured experimentally~\cite{tinkham2004introduction}. One can see in Fig.~\ref{fig:3}~(b) that the peak in absorption is much larger than the absorption of the irradiation in absence of a condensate flow. Thus, it can be easily measured in experiments.

Note also that the ac admittance of Al samples with different concentration of impurities has been measured at ${T = 0.355 T_{\text{c}}}$ in an applied magnetic field, i.e., in the presence of a dc supercurrent, by Budzinski~\emph{et al.}~\cite{Budzinski_Garfunkel_Markley_1973}. A peak in the absorption near the frequency~$2 \Delta$ has been observed in samples with sufficiently high impurity concentration and magnetic field. The effect predicted here may serve as explanation of the obtained experimental results and to the best of our knowledge, there is no other satisfactory theory of this experiment.

In conclusion, we have analyzed the excitation of the amplitude mode in superconductors by a weak ac~irradiation in the presence of a supercurrent~$I_{0}$. We have shown that the condensate flow leads to a coupling of electromagnetic field to the modulus of the order parameter so that the AM can be excited even in the linear approximation in amplitude of the ac electric field ${\mathbf{E}_{\Omega} = (-i \Omega / e) \hbar \mathbf{Q}_{\Omega}}$. The amplitude of the perturbation of the superconducting OP is proportional to the scalar product of the electric field and the velocity of the condensate, ${\delta \Delta_{\Omega} \propto \mathbf{Q}_{\Omega} \mathbf{Q}_{0}}$. The intensity of the signal depends on the polarization of the incident electric field and has a resonance at ${\Omega = 2 \Delta}$. These features enable a simple identification of the Higgs mode in superconductors by measuring the admittance with a linearly polarized light. We emphasize that our method probes the same Higgs mode as the one measured in the recent experiments Refs.~\onlinecite{ExpS13,Matsunaga_et_al_2014}. Of course, one could measure the admittance on the same setups as those employed in these experiments just using one (even weak) laser pulse. The transmission or reflection coefficients of this pulse have a peak, respectively, dip at ${\Omega = 2 \Delta}$. The important feature of the mechanism of the AM excitation by a weak electromagnetic field is that it acts directly on the order parameter~$\Delta$ not perturbing other order parameters (for example, charge density wave) which can coexist with~$\Delta$, e.g., in high\nobreakdash-$T_{\text{c}}$ superconductors. Combining this and conventional two-photon absorption methods for studying the AM, one can obtain important information about dynamics of different order parameters~\cite{Moor14,Levchenko15}.

\newpage

\appendix

\section{Supplemental Material}
\label{SM}

\subsection{Variation of the order parameter~$\delta \hat{\Delta}$.}

We assume that a superconducting current flowing in a superconducting film consists of a constant and oscillating parts, that is, the condensate momentum is a sum of two terms,
\begin{equation}
\mathbf{Q}(t) = \mathbf{Q}_{0} + \mathbf{Q}_{\Omega} \exp (i \Omega t) \,. \label{S1}
\end{equation}

We propose that both components are small, i.e., ${D Q_{0,\Omega}^{2} \ll \Delta}$. Our task is to find the response of the superconductor to~$\mathbf{Q}_{0,\Omega}$, i.e., to find~$\check{g}_{\text{st}}$ and~$\delta \check{g}_{\Omega}$ from Eq.~(15) of the main text. Since the superconductor remains in equilibrium in the presence of a constant condensate flow~$\mathbf{Q}_{0}$, the elements of the matrix~$\check{g}_{\text{st}}$ have equilibrium forms, that is, diagonal elements are~$\hat{g}_{\text{st}}^{R}$, $\hat{g}_{\text{st}}^{A}$ and the off-diagonal element (Keldysh function) is
\begin{equation}
\check{g}|_{12} \equiv \hat{g}_{\text{st}}^{K} = (\hat{g}_{\text{st}}^{R} - \hat{g}_{\text{st}}^{A}) \tanh (\epsilon \beta) \,, \label{S2}
\end{equation}
where ${\beta = 1 / (2T)}$. The functions ${\hat{g}_{\text{st}}^{R(A)} \simeq \hat{g}_{0}^{R(A)} + \delta \hat{g}_{0}^{R(A)}}$, where ${\hat{g}_{0}^{R(A)} = g_0^{R(A)} \hat{\tau}_{3} + i \hat{\tau}_{2} f_0^{R(A)}}$ have the well known form
\begin{align}
g_{0}^{R(A)}(\epsilon) &= \epsilon / \zeta_{0}^{R(A)}(\epsilon) \,, \label{S3} \\
f_{0}^{R(A)}(\epsilon) &= \Delta_{0} / \zeta_{0}^{R(A)}(\epsilon) \,, \label{S3'}
\end{align}
with ${\zeta_{0}^{R(A)}(\epsilon) = \sqrt{(\epsilon \pm i \gamma )^{2} - \Delta_{0}^{2}}}$. The corrections~$\delta \hat{g}_{0}^{R(A)}$ can be easily obtained from the linearized equation~(15) and normalization condition~(13) of the main text which has the form
\begin{equation}
\big[ \delta \hat{g}_{0} \cdot \hat{g}_{0} + \hat{g}_{0} \cdot \delta \hat{g}_{0} \big]^{R(A)} = 0 \,. \label{S4'}
\end{equation}

We find for $\delta \hat{g}_{0}^{R(A)}$
\begin{equation}
\delta \hat{g}_{0}^{R(A)}(\epsilon) = \Bigg[\frac{\delta \hat{\Delta} - (\hat{g}_{0} \delta \hat{\Delta} \hat{g}_{0}) + i W_{Q}(\hat{\bar{g}}_{0} - \hat{g}_{0} \hat{\bar{g}}_{0} \hat{g}_{0})}{2 \zeta_{0}} \Bigg]^{R(A)} \,, \label{S4}
\end{equation}
where ${W_{Q} = D Q_{0}^{2}}$. The correction ${\delta \hat{\Delta}_{0} = \delta \Delta_{0} i \hat{\tau}_{2}}$ has to be determined from the self-consistency equation
\begin{equation}
\delta \Delta_{0} = -i \lambda \mathrm{Tr} \hat{\tau}_{2} \int d \epsilon \big[ \delta \hat{g}_{0}^{R}(\epsilon) - \delta \hat{g}_{0}^{A}(\epsilon) \big] \tanh(\epsilon \beta) \,. \label{S5}
\end{equation}

Substituting Eq.~(\ref{S4}) into Eq.~(\ref{S5}), we find
\begin{equation}
\delta \Delta_{0} = - \frac{2W_{Q}}{\Delta} \frac{\sum_{\omega} \omega^{2} \zeta_{\omega}^{-4}}{\sum_{\omega} \zeta_{\omega}^{-3}} \,, \label{S6}
\end{equation}
where ${\omega = (2n+1) \pi T}$ is the Matsubara frequency and ${\zeta_{\omega}^{2} = (\omega^{2} + \Delta^{2})}$.

Now, we find the correction~$\delta \check{g}_{\Omega}$ caused by an external irradiation with frequency~$\Omega$. We represent~$\delta \hat{g}_{\Omega}^{K}$ as a sum of a regular,~$\delta \hat{g}_{\Omega}^{\text{reg}}$ and anomalous,~$\hat{g}_{\Omega}^{\text{an}}$ parts~\cite{G-Eliash,ArtVolkovRev80}
\begin{equation}
\delta \hat{g}_{\Omega}(\epsilon_{+},\epsilon_{-}) = \delta \hat{g}_{\Omega}^{\text{reg}}(\epsilon_{+},\epsilon_{-}) + \hat{g}_{\Omega}^{\text{an}}(\epsilon_{+},\epsilon_{-}) \,, \label{S7}
\end{equation}
with
\begin{equation}
\delta \hat{g}_{\Omega }^{\text{reg}}(\epsilon_{+},\epsilon_{-}) = \delta \hat{g}_{\Omega}^{R}(\epsilon_{+},\epsilon_{-}) \tanh (\epsilon_{-} \beta) - \tanh(\epsilon_{+} \beta) \delta \hat{g}_{\Omega}^{A}(\epsilon_{+},\epsilon_{-}) \,, \label{S8}
\end{equation}
where ${\epsilon_{\pm} = \bar{\epsilon} \pm \Omega / 2}$ and ${\bar{\epsilon} = (\epsilon + \epsilon^{\prime}) / 2}$. In order to determine the regular part~$\delta \hat{g}_{\Omega}^{\text{reg}}(\epsilon_{+},\epsilon_{-})$, we need to find the corrections~$\delta \hat{g}_{\Omega}^{R(A)}(\epsilon_{+},\epsilon_{-})$. These corrections can be easily found from the linearized equation~(15) and normalization condition (see main text),
\begin{equation}
\delta \hat{g}^{R(A)} \cdot \hat{g}_{-}^{R(A)} + \hat{g}_{+}^{R(A)} \cdot \delta \hat{g}^{R(A)} = 0 \,, \label{S8'}
\end{equation}%
as it was done in Ref.~\cite{ArtVolkovRev80}. The result is
\begin{equation}
\delta \hat{g}_{\Omega}^{R} = \frac{(\delta \hat{\Delta} - \hat{g}_{+}^{R} \delta \hat{\Delta} \hat{g}_{-}^{R}) - i \delta W_{Q} \hat{m}^{R}}{\zeta_{+}^{R} + \zeta_{-}^{R}} \,, \label{S9}
\end{equation}
where ${\delta W_{Q} = D(\mathbf{Q}_{0} \mathbf{Q}_{\Omega}) = D Q_{0} Q_{\Omega} \cos \theta}$, with~$\theta$ being the angle between the vectors~$\mathbf{Q}_{0}$ and~$\mathbf{Q}_{\Omega}$, ${\hat{g}_{\pm}^{R} \equiv \hat{g}_{0}^{R}(\epsilon_{\pm})}$, ${\hat{m}^{R} = \big[\hat{g}_{+} \cdot (\hat{\bar{g}}_{-} + \hat{\bar{g}}_{+}) \cdot \hat{g}_{-} - (\hat{\bar{g}}_{-} + \hat{\bar{g}}_{+})\big]^{R}}$, ${\hat{\bar{g}}_{\pm} \equiv (\hat{\tau}_{3} \cdot \hat{g}_{\pm}^{R} \cdot \hat{\tau}_{3})}$, and ${\zeta_{\pm}^{R(A)} = \sqrt{(\epsilon _{\pm} \pm i \gamma)^{2} - \Delta_{0}^{2}}}$. The same formula takes place for~$\delta \hat{g}_{\Omega}^{A}$, and a similar formula can be obtained for the anomalous part~$\hat{g}_{\Omega}^{\text{an}}$,
\begin{equation}
\hat{g}_{\Omega}^{\text{an}} = \frac{\big[ \tan (\epsilon _{+} \beta) - \tan (\epsilon_{-} \beta) \big] \big[(\delta \hat{\Delta} - \hat{g}_{+}^{R} \delta \hat{\Delta} \hat{g}_{-}^{A}) - i \delta W_{Q} \hat{m}^{\text{an}} \big]}{\zeta_{+}^{R} + \zeta_{-}^{A}} \,,  \label{S10}
\end{equation}%
where~$\hat{m}^{\text{an}}$ coincides with~$\hat{m}^{R}$ if the matrix~$\hat{g}_{-}^{R}$ in the expression for~$\hat{m}^{R}$ is replaced by~$\hat{g}_{-}^{A}$.

Knowing the matrix functions~$\delta \hat{g}_{\Omega }^{\text{reg}}(\epsilon_{+}, \epsilon_{-})$ and~$\hat{g}_{\Omega}^{\text{an}}(\epsilon_{+}, \epsilon_{-})$ we can determine the Fourier components of the variations of the order parameter~$\delta \Delta_{\Omega}$ and of the current~$\delta j_{\Omega}$. The former is determined from the self-consistency equation which is a generalization of Eq.~(\ref{S5}) for a nonstationary case,
\begin{equation}
\delta \Delta_{\Omega} = -i \lambda \mathrm{Tr} \hat{\tau}_{2} \int_{-E_{\text{D}}}^{E_{\text{D}}} d \bar{\epsilon} \big[\delta \hat{g}^{\text{reg}}(\epsilon_{+},\epsilon_{-}) + \hat{g}^{\text{an}}(\epsilon_{+},\epsilon_{-}) \big] \,, \label{S11}
\end{equation}
where~$E_{\text{D}}$ is the Debye energy and the matrices~$\delta \hat{g}^{\text{reg}}(\epsilon_{+},\epsilon_{-})$ and~$\hat{g}^{\text{an}}(\epsilon_{+},\epsilon_{-})$ are given by Eqs.~(\ref{S9}) and~(\ref{S10}). It is useful to consider also the identity
\begin{equation}
\delta \Delta_{\Omega} = -i \lambda \delta \Delta_{\Omega} \mathrm{Tr} \hat{\tau}_{2} \int_{-E_{\text{D}}}^{E_{\text{D}}} d \bar{\epsilon} \big[ \frac{1}{\zeta^{R}(\bar{\epsilon})} - \frac{1}{\zeta^{A}(\bar{\epsilon})} \big] \tanh (\bar{\epsilon} \beta) \,. \label{S12}
\end{equation}
Subtracting Eq.~(\ref{S12}) from Eq.~(\ref{S11}) we obtain
\begin{equation}
\delta \Delta_{\Omega} \big[ A^{\text{reg}} + A^{\text{an}} \big] = i \delta W_{Q} \big[ B^{\text{reg}} + B^{\text{an}} \big] \,.\label{S13}
\end{equation}
Here, the functions~$A^{\text{reg}/\text{an}}$ and~$B^{\text{reg}/\text{an}}$ are
\begin{widetext}
\begin{align}
A^{\text{reg}} &= \int \Bigg[ \frac{\zeta_{+}^{R} \zeta_{-}^{R} + \epsilon_{+} \epsilon_{-} + \Delta^{2}}{\zeta_{+}^{R} \zeta_{-}^{R} \big( \zeta_{+}^{R} + \zeta_{-}^{R} \big)} \tanh (\epsilon_{-} \beta) - \frac{\tanh (\bar{\epsilon} \beta)}{\zeta^{R}(\bar{\epsilon})} \Bigg] - \Bigg[ \frac{\zeta_{+}^{A} \zeta_{-}^{A} + \epsilon_{+} \epsilon_{-} + \Delta^{2}}{\zeta_{+}^{A} \zeta_{-}^{A} \big( \zeta_{+}^{A} + \zeta_{-}^{A} \big)} \tanh (\epsilon_{+} \beta) - \frac{\tanh (\bar{\epsilon} \beta)}{\zeta^{A}(\bar{\epsilon})} \Bigg] \, \mathrm{d} \bar{\epsilon} \,, \\
A^{\text{an}} &= \int \frac{\zeta_{+}^{R} \zeta_{-}^{A} + \epsilon_{+} \epsilon_{-} + \Delta^{2}}{\zeta_{+}^{R} \zeta_{-}^{A} \big( \zeta_{+}^{R} + \zeta_{-}^{A} \big)} \big[ \tanh (\epsilon_{+} \beta) - \tanh (\epsilon_{-} \beta) \big] \, \mathrm{d} \bar{\epsilon} \label{S14} \,, \\
B^{\text{reg}} &= 2 \Delta \int \Bigg[ \frac{\epsilon_{+} \zeta_{-}^{R} + \epsilon_{-} \zeta_{+}^{R}}{\big( \zeta_{+}^{R} \zeta_{-}^{R} \big)^{2} \big( \zeta_{+}^{R} + \zeta_{-}^{R} \big)} \big( \epsilon_{+} + \epsilon_{-} \big) \tanh (\epsilon_{-} \beta) \Bigg] - \Bigg[ \frac{\epsilon_{+} \zeta_{-}^{A} + \epsilon_{-} \zeta_{+}^{A}}{\big( \zeta_{+}^{A} \zeta_{-}^{A} \big)^{2} \big( \zeta_{+}^{A} + \zeta_{-}^{A} \big)} \big( \epsilon_{+} + \epsilon_{-} \big) \tanh (\epsilon_{+} \beta) \Bigg] \, \mathrm{d} \bar{\epsilon} \,, \label{18'} \\
B^{\text{an}} &= 2 \Delta \int \frac{\epsilon_{+} \zeta_{-}^{A} + \epsilon_{-} \zeta_{+}^{R}}{\big( \zeta_{+}^{R} \zeta_{-}^{A} \big)^{2} \big( \zeta_{+}^{R} + \zeta_{-}^{A} \big)} \big( \epsilon_{+} + \epsilon_{-} \big) \big[ \tanh (\epsilon_{+} \beta) - \tanh (\epsilon_{-} \beta) \big] \, \mathrm{d} \bar{\epsilon} \,, \label{S15}
\end{align}
\end{widetext}

\subsection{Ac current.} The ac electric current is given by the formula
\begin{widetext}
\begin{align}
\mathbf{I}_{\Omega} &= (\sigma / 4) \mathrm{Tr} \big\{ \hat{\tau}_{3} \cdot [\hat{g}(t) \cdot \nabla \hat{g}(t)]^{K} \big\}|_{\Omega} \\
                    &= i (\sigma / 4) \int d \bar{\epsilon} \big[ \mathbf{Q}_{\Omega} \mathrm{Tr} \big\{ [\hat{g}_{-} \cdot \hat{\bar{g}}_{+} - 1]^{K} \big\} + \mathbf{Q}_{0} \mathrm{Tr} \big\{ [\delta \hat{g}_{-} \cdot \hat{\bar{g}} + \hat{\bar{g}}_{+} \cdot \delta \hat{g}]^{K} \big\} \big] \,.\label{A1}
\end{align}
\end{widetext}

The current~$I_{\Omega}$ can be written in the form
\begin{equation}
\mathbf{I}_{\Omega} = \mathbf{I}^{(1)} + \mathbf{I}^{(2)} \,, \label{A2}
\end{equation}
where~$\mathbf{I}^{(1)}$ equals the first term and~$\mathbf{I}^{(2)}$ equals the second term in Eq.~(\ref{A2}). Note an important point. The currents~$I^{(1)}$ and~$I^{(2)}$ can be measured separately because the current~$I^{(1)}$ does not depend on the angle~$\theta$ between the vectors~$\mathbf{Q}_{0}$ and~$\mathbf{Q}_{\Omega}$, whereas the current~${I^{(2)} \propto Q_{0} \delta W_{Q} \propto Q_{0} \cos \theta}$, see Eq.~(\ref{S9}).

First, we consider the current~$\mathbf{I}^{(1)}$. Since~$\mathbf{Q}_{0}$ is assumed to be small, one can write~$I^{(1)}$ as
\begin{equation}
\mathbf{I}^{(1)} = \mathbf{I}_{0}^{(1)} + \delta \mathbf{I}^{(1)} \,, \label{A3}
\end{equation}
where
\begin{widetext}
\begin{equation}
\mathbf{I}_{0}^{(1)} = i (\sigma / 4) \mathbf{Q}_{\Omega} \int d \bar{\epsilon} \big[ j_{0}^{R} \tanh(\epsilon_{-} \beta) - j_{0}^{A} \tanh(\epsilon_{+} \beta) + j_{0}^{\text{an}}[\tanh(\epsilon_{+} \beta) - \tanh(\epsilon_{-} \beta)] \big]
\label{A3a}
\end{equation}
\end{widetext}
with
\begin{align}
j_{0}^{R(A)} &= \big[ g_{+} g_{-} + f_{+} f_{-} \big]^{R(A)} - 1 \,, \\
j_{0}^{\text{an}} &= \big[ g_{+}^{R} g_{-}^{A} + f_{+}^{R} f_{-}^{A} \big] - 1 \,. \label{A3'}
\end{align}
This current at ${Q_{0} = 0}$ has been calculated by Abrikosov and Gor'kov~\cite{AbrGor58} by another method long ago. In a general case of a nonzero ${Q_{0} \neq 0}$, the Green's functions~$\hat{g}_{Q}(\epsilon)$ depend on~$Q_{0}$. The current~$\mathbf{I}^{(1)}$ for low~$\Omega$ has been calculated in Ref.~\onlinecite{Clem_Kogan_2012} for arbitrary~$Q_{0}$.

The correction~$\delta \mathbf{I}^{(1)}$ to the linear response due to moving condensate can be easily found with the help of corrections to the Green's functions~$\delta \hat{g}_{0}^{R(A)}(\epsilon)$,~Eq.~(\ref{S4}). We obtain an equation similar to Eq.~(\ref{A3a}),
\begin{widetext}
\begin{equation}
\delta \mathbf{I}^{(1)} = i (\sigma / 4) \mathbf{Q}_{\Omega} \int d \bar{\epsilon} \big[ j_{1}^{R} \tanh(\epsilon_{-} \beta) - j_{1}^{A} \tanh(\epsilon_{+} \beta) + j_{1}^{\text{an}} [\tanh(\epsilon_{+} \beta) - \tanh(\epsilon_{-} \beta)] \big] \label{A4}
\end{equation}
\end{widetext}
with~$j_{1}^{R(A)}$ and~$j_{1}^{\text{an}}$ given by
\begin{align}
j_{1}^{R(A)} &= N^{R} \Bigg[ \delta \Delta_{0} \Big( \frac{\epsilon_{+}}{\zeta_{+}^{2}} + \frac{\epsilon_{-}}{\zeta_{-}^{2}} \Big) - 2 i W_{Q} \Delta \Big( \frac{\epsilon_{+}}{\zeta_{+}^{3}} + \frac{\epsilon_{-}}{\zeta_{-}^{3}} \Big) \Bigg]^{R(A)} \,, \label{A4a} \\
j_{1}^{\text{an}} &= N^{\text{an}} \Bigg[ \delta \Delta_{0} \Big( \frac{\epsilon_{+}}{(\zeta_{+}^{R})^{2}} + \frac{\epsilon_{-}}{(\zeta_{-}^{A})^{2}} \Big) - 2 i W_{Q} \Delta \Big( \frac{\epsilon_{+}}{(\zeta_{+}^{R})^{3}} + \frac{\epsilon_{-}}{(\zeta_{-}^{A})^{3}} \Big) \Bigg] \,, \label{A4b}
\end{align}
where ${N^{R(A)} = \big[ g_{+} f_{-} + g_{-} f_{+} \big]^{R(A)}}$, ${N^{\text{an}} = g_{+}^{R} f_{-}^{A} + g_{-}^{A} f_{+}^{R}}$, and $\delta \Delta_{0}$ is provided in Eq.~(\ref{S6}).

The current~${\delta \mathbf{I}^{(2)} = \mathbf{I}^{(2)} - \mathbf{I}^{(2)}_0}$ is presented in the form similar to~Eq.~(\ref{A4}),
\begin{widetext}
\begin{equation}
\delta \mathbf{I}^{(2)} = i (\sigma / 4) \mathbf{Q}_{0} \int d \bar{\epsilon} \big[ j_{2}^{R} \tanh(\epsilon_{-} \beta) - j_{2}^{A} \tanh(\epsilon_{+} \beta) + j_{2}^{\text{an}} [\tanh(\epsilon_{+} \beta) - \tanh(\epsilon_{-} \beta) ] \big] \,,
\label{A6}
\end{equation}
\end{widetext}
where the terms~$j_{2}^{R(A)}$, $j_{2}^{\text{an}}$ are found with the help of Eqs.~(\ref{S9}) and~(\ref{S10}). They are equal to
\begin{widetext}
\begin{align}
j_{2}^{R(A)} &= \frac{\delta \Delta_{\Omega} \big[ F (1 + M) + G N \big]^{R(A)} + i \delta W_{Q} \big[G^{2} (1 - M) - F^{2} (1 + M) - 2 F G N \big]^{R(A)}}{\big(\zeta_{+} + \zeta_{-} \big)^{R(A)}} \,, \\
j_{2}^{\text{an}} &= \frac{\delta \Delta_{\Omega} \big[ F (1 + M) + G N \big]^{\text{an}} + i \delta W_{Q} \big[ G^{2} (1 - M) - F^{2} (1 + M) - 2 F G N \big]^{\text{an}}}{\big(\zeta_{+}^{R} + \zeta_{-}^{A} \big)} \,, \label{A7}
\end{align}
\end{widetext}
where
\begin{align}
M^{R(A)} &= \big[ g_{+} g_{-} + f_{+} f_{-} \big]^{R(A)} \,, & M^{\text{an}} &= g_{+}^{R} g_{-}^{A} + f_{+}^{R} f_{-}^{A} \,, \\
N^{R(A)} &= \big[ g_{+} f_{-} + g_{-} f_{+} \big]^{R(A)} \,, & N^{\text{an}} &= g_{+}^{R} f_{-}^{A} + g_{-}^{A} f_{+}^{R} \,, \\
G^{R(A)} &= \big[ g_{+} + g_{-} \big]^{R(A)} \,, & G^{\text{an}} &= \big[ g_{+}^{R} + g_{-}^{A} \big]^{\text{an}} \,, \\
F^{R(A)} &= \big[ f_{+} + f_{-} \big]^{R(A)} \,, & F^{\text{an}} &= f_{+}^{R} + f_{-}^{A} \,.
\end{align}

Using the expressions for~$g_{\pm}^{R(A)}$ and~$f_{\pm}^{R(A)}$, i.e., ${g_{\pm}^{R(A)} = (\bar{\epsilon} \pm \Omega / 2) / \zeta_{\pm}^{R(A)} = f_{\pm}^{R(A)} \Delta / (\bar{\epsilon} \pm \Omega / 2)}$, one can calculate (numerically in a general case) the admittances ${Y^{(1)} = \delta I^{(1)} / E_{\Omega}}$ and~${Y^{(2)} = \delta I^{(2)} (\cos \theta) / E_{\Omega}}$, where ${E_{\Omega} = -i \Omega Q_{\Omega} / e}$ and we emphasized the dependence of the current~$\delta I^{(2)}$ on the angle between the vectors~$\mathbf{Q}_{0}$ and~$\mathbf{Q}_{\Omega}$. If we write the currents~$\delta I^{(1,2)}$ in the form ${\delta I^{(1,2)} = (K^{\prime} + i K^{\prime \prime}) Q_{\Omega}}$, then the real and imaginary parts of the admittance~$Y^{\prime}$ and~$Y^{\prime \prime}$ are related to~$K^{\prime}$ and~$K^{\prime \prime}$, respectively, via the expressions
\begin{align}
Y^{\prime} = - K^{\prime \prime} / \Omega \,, \\
Y^{\prime \prime} = -K^{\prime} / \Omega \,. \label{A8}
\end{align}

The admittance can be written in the form similar to Eq.~(\ref{A2}),
\begin{equation}
Y_{\Omega} = Y_{0 \Omega}^{(1)} + \delta Y_{\Omega}^{(1)} + Y_{\Omega}^{(2)} \,, \label{A9}
\end{equation}
where the first term~$Y_{0\Omega}^{(1)}$ is the admittance of a superconductor with ${Q_{0} = 0}$ calculated in Refs.~\onlinecite{Mattis_Bardeen_1958,AbrGor58}. The second term is a correction to~$Y_{0\Omega}^{(1)}$ due to steady motion of the condensate and the third term, ${Y_{\Omega}^{(2)} \propto \delta \Delta_{\Omega} \propto \mathbf{Q}_{0} \cdot \mathbf{Q}_{\Omega}}$, depends on the mutual polarization of the vectors~$\mathbf{Q}_{0}$ and~$\mathbf{Q}_{\Omega}$. It is proportional to the amplitude mode~$\delta \Delta_{\Omega}$. One can easily express~$\delta Y_{\Omega}^{(1)}$ and~$Y_{\Omega}^{(2)}$ in terms of coefficients in Eq.~(9) of the main text,
\begin{align}
\delta Y_{\Omega}^{(1)} &= i \frac{e}{\Omega} \frac{\partial K(Q_{0},\Omega)}{\partial Q_{0}^{2}} Q_{0}^{2} \,, \\
Y_{\Omega}^{(2)} &= i \frac{e}{\Omega} \varkappa_{\text{res}}(\Omega) Q_{0}^{2} \cos^{2} \vartheta \,.  \label{A10}
\end{align}
Equation~(\ref{A10}) represents, in fact, the definition of the function~$\varkappa_{\text{res}}$.

The admittance~$Y(\Omega)$ is connected with dielectric penetrability ${\varepsilon(\Omega) \equiv \varepsilon_{1}(\Omega) + i \varepsilon_{2}(\Omega)}$, ${Y(\Omega) = - i \Omega \varepsilon(\Omega)}$, and the latter quantity determines the reflection coefficient~$R$~\cite{L-L8},
\begin{equation}
R = \Bigg| \frac{\sqrt{\varepsilon_{1}(\Omega)} - \sqrt{\varepsilon_{2}(\Omega)}}{\sqrt{\varepsilon_{1}(\Omega)} + \sqrt{\varepsilon_{2}(\Omega)}} \Bigg|^{2} \,,
\end{equation}
which is usually measured in optical experiments.


%

\end{document}